\documentclass{iau}
\usepackage{graphicx}

\title[IAUS~342~~XRISM and Athena] 
{The Hot Universe with XRISM and Athena}

\author[Matteo Guainazzi \& Makoto~S.Tashiro]   
{Matteo Guainazzi$^1$ \and Makoto~S. Tashiro$^{2,3}$}

\affiliation{$^1$ESTEC/ESA, Keplerlaan 1, 2201AZ Noordwijk, The Netherlands \\ email: {\tt Matteo.Guainazzi@sciops.esa.int} \\[\affilskip]
$^2$ISAS/JAXA, 3-1-1 Yoshinodai, Chuo-ku, Sagamihara, Kanazgawa 252-5210, Japan \\email: {\tt tashiro@astro.isas.jaxa.jp} \\[\affilskip]
$^3$Graduate School of Science and Engineering, Saitama University, 255 Shimo-Okubo, Sakawa, Saitama, Saitama 338-8570, Japan \\email: {\tt tashiro@phy.saitama-u.ac.jp}}

\pubyear{2019}
\volume{IAUS342}  
\jname{Perseus in Sicily: from black hole to cluster outskirts}
\editors{K.~Asada, E~de Gouveia dal Pino, H.~Nagai, \\ R.~Nemmen, \& M.~Giroletti eds.}
\begin{document}

\maketitle

\begin{abstract}

  X-ray spectroscopy is key to address the theme of ``The Hot Universe'', the
  still poorly understood astrophysical processes driving the cosmological evolution
  of the baryonic hot gas traceable through its electromagnetic radiation.
  Two future X-ray observatories: the JAXA-led XRISM (due to launch
  in the early 2020s),
  and the ESA Cosmic Vision L-class mission {\it Athena} (early 2030s)
  will provide breakthroughs in our understanding of how and when large-scale
  hot gas structures formed in the Universe, and in tracking their
  evolution from the formation epoch to the present day.
  
\keywords{space vehicles: instruments, telescopes, X-rays: general}

\end{abstract}

\firstsection 

\section{The Hot Universe}

 The last a few decades
have witnessed a momentous improvement in our understanding of the
origin and evolution of the Universe.
According to the standard $\Lambda$CDM cosmology, the Big Bang was followed by
a short (a fraction of a second)
phase of strongly accelerated expansion (``inflation''),
followed by a phase of slower but continuing expansion.
Density fluctuations in the early Universe led to the formation of the
first structures. These seeds grew
over cosmic time via hierarchical collapse. While the
evolution of large scale structures is primarily driven by the cosmological
parameters and by the interaction with the gravitationally dominant dark matter,
still
poorly understood astrophysical processes drive the evolution of the {\bf ``Hot
Universe''}, the baryonic hot gas
traceable through its electromagnetic radiation. Accretion
onto the dark matter potential wells heated gas to million degree
temperatures. Galaxy
clusters (the largest gravitationally bound structures in the Universe)
lay at the nodes of the cosmic web
(\cite{schaye15}). While about 80\% of their total mass
is in the form of dark matter, the remainder is dominated by diffuse, hot,
metal-enriched X-ray emitting plasma (the ``Intergalactic Cluster Medium, ICM),
with stars constituting only 15\% of visible baryons.  Furthermore, it is now ascertained
that a Super-Massive Black Hole (SMBH) sits at the center of
most, if not any, galaxies. SMBHs
can inject sufficient energy into the ICM to substantially affect its
structure, dynamical state and chemical abundance
(\cite{fabian12}). This ``feedback'' process can
be augmented by SuperNova (SN) winds, coupling the ICM evolution with
the life cycle of stars in galaxies. The evolution of the ICM can be therefore
shaped by processes beyond the pure gravitational
collapse. By tracing clusters and groups from the local Universe back to
their formation epochs (at z$\sim$2--3), observational cosmology can
be put on a solid experimental grounds because a full understanding of
the entire range of astrophysical process at play is required to get a
consistent picture of the evolution of cosmic structures. Last,
but not least, the census of baryons is largely incomplete,
$\simeq$40\% at z$<$2 still eluding
detection despite intense observational efforts
(\cite{nicastro17}; see, however, \cite{nicastro18}).

Major astrophysical questions remain therefore to answer (\cite{nandra13}):

\begin{itemize}

\item How do baryons in groups and clusters accrete and dynamically evolve
  in the dark matter halos?

\item What drives the chemical and thermodynamic evolution of the Universe's
  largest structures?

\item What is the interplay of galaxies, super-massive black holes, and
  intergalactic gas evolution in groups and clusters?

\item Where are the missing baryons at low redshift, and what is their physical
  states?
  
\end{itemize}

X-ray is the only band of the electromagnetic
spectrum allowing to probe physical properties of this plasma
such as temperature, densities,
abundances, velocity field, and ionisation state. All
these quantities can be reliably measured only with high-resolution
X-ray spectroscopic observations able to characterize
recombination transitions
of highly ionized (He- and H-like) ions of a wide range of metals.
Such measurements allow us to fully characterize the physical state
of the plasma as well as the relative share of the energy going into
thermal and non-thermal processes in the ICM, the latter
generating turbulence and
kpc-scale bulk motions. These processes have still to be observed at
the relevant spatial scales (corresponding to a few arc-seconds in the
local Universe). Furthermore, comparing abundances in the local and
high-redshift Universe will trace the evolution of the chemical enrichment
and the impact of SN feedback on the growth of large-structure in the
Universe. Comparing abundances in galaxies, galaxy groups and
clusters and in the intergalactic medium is needed to
model the life-cycle of
matter. All these measurements require a significant improvement in
the performance of X-ray spectroscopic instruments over several parameter
spaces: energy resolution, spatial resolution, and collecting area.

\section{The {\it Hitomi} heritage}

However, high-resolution X-ray spectroscopy is just in its infancy.
Only with the advent of modern X-ray observatories such as {\it Chandra}
and XMM-Newton at the beginning of the century, X-ray detectors with
a resolving power higher than 100 became routinely operational
(\cite{paerels03}). These
missions, still enormously successful,
carry grating systems able to disperse the incoming X-ray photons
onto energy-sensitive detectors. In the soft
X-ray band (E$\le$2~keV) resolving powers of $\sim$3--400, and $\sim$1000
were
achieved by the grating systems on board XMM-Newton
(\cite{denherder01}) and {\it Chandra} (\cite{brinkman87,canizares05}),
respectively.
In the
energy range around $\simeq$6~keV, where the all-important K-shell
transitions of iron occur, the {\it Chandra} high-energy gratings
achieve a resolving power of $\simeq$170, albeit with a tiny effective
area ($\simeq$30~cm$^2$).

\begin{figure}[b]
\begin{center}
 \includegraphics[width=5in]{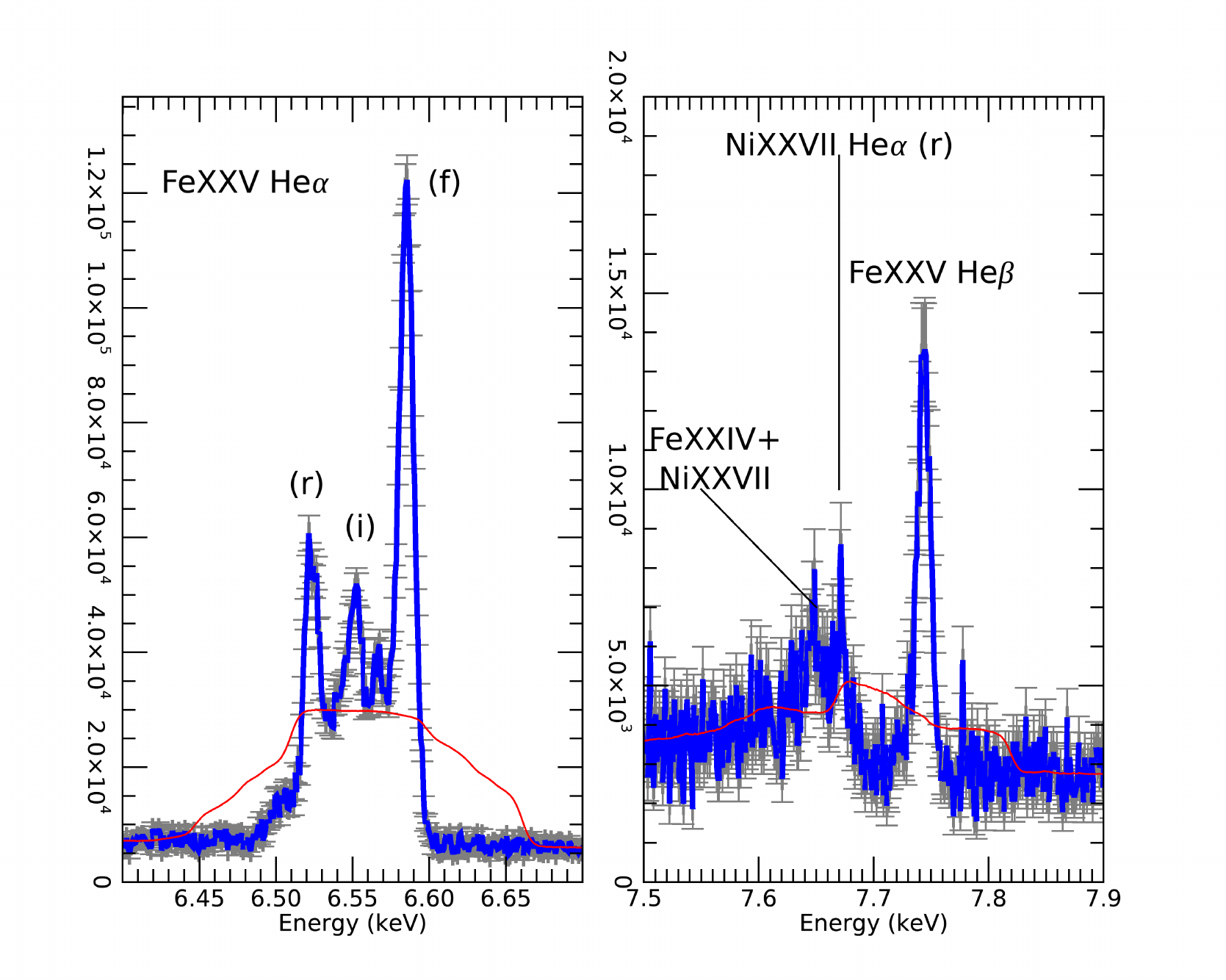} 
 \caption{{\it Hitomi}/SXS spectrum of the Perseus Cluster in the 6.4--6.7~keV
   ({\it left panel})
   and 7.5--7.9~keV energy range ({\it right panel}), respectively. The
   main emission lines discussed in the manuscript are labeled. The error bars
   correspond to 1$\sigma$ Poissonian uncertainties. The {\it red lines}
   represent the same spectrum at CCD-resolution.
   Technical details:
   the spectrum was extracted from the combined calibrated event lists
   of Obs.\# 100040020, -30, -40, and -50
   ($\simeq$2.43$\times$10$^5$~s net exposure
   time), downloaded from the {\it Hitomi} science archive. It
   corresponds to a circular extraction region around the
   cluster core with a 150'' radius. It was generated with {\sc HEASOFT}
   version 6.22, and the calibration files version 10, released on 15 February
 2018. }
   \label{fig1}
\end{center}
\end{figure}

A huge step forward was expected with the JAXA-led X-ray observatory {\it
Hitomi}, whose successful launch on 2016 February 17 led to hope that
a new era in observational X-ray spectroscopy had started. {\it Hitomi} carried the
{\it Soft X-ray Spectrometer} (SXS),
a pixellated micro-calorimeter detector with an
unprecedented $\le$5~eV energy resolution over the 0.3--12~keV energy band
(\cite{kelley16}). Such a performance
corresponds to an improvement in resolving power
larger then one order of magnitude at 6~keV with respect to the {\it Chandra}
high-energy transmission gratings, together with a one-order-of magnitude larger
effective area at the same energies. Regrettably,
the spacecraft was lost after only six
weeks of operations due to a chain of anomalies of the attitude control system
coupled to human errors (\cite{jaxa16}).
However, careful planning during the commissioning phase
led to {\it Hitomi}
observing for almost one week the core of the Perseus galaxy
cluster, one of the X-ray brightest in the local Universe
($z$~=~0.0179). While the SXS was still not formally
commissioned, it was still operated with the closed ``gate valve''
with a $\sim$300~$\mu$m beryllium window along the optical
path as contamination prevention (leading to the total suppression of photons
at energies $\le$2~keV), and its data
were largely self-calibrated (\cite{tsujimoto18}), its spectra provided a
transformational view of the properties of the ICM in Perseus
(Fig.~\ref{fig1}).

The first surprise came from the analysis of the dynamical state of the ICM.
Based primarily on the Fe~{\sc XXV} multiplet
(see the {\it left panel} of Fig.~\ref{fig1}),
the SXS measured a line-of-sight velocity dispersion of the
turbulent gas of 164$\pm$10~km~s$^{-1}$, in a region 30--60~kpc from the
central nucleus, with only a
marginally larger velocity in the core (187$\pm$13~km~s$^{-1}$)
(\cite{hitomi16}). This level of precision is unprecedented in X-ray
astronomy. The measured velocities correspond to a turbulent pressure
$\simeq$4\% ($<$8\%) of the thermal pressure. This is puzzling,
because a powerful radio-loud Active Galactic Nucleus (AGN;
NGC1275)
is present at the cluster core, injecting an amount of
energy in the ICM sufficient to
evacuate bubbles filled with relativistic plasma
(\cite{boehringer93,fabian00}): one of the most spectacular examples of
``radio-mode'' AGN feedback, postulated to prevent radiative cooling of the
ICM core gas. The {\it Hitomi} result is
important not only for its implications on the astrophysics of the ICM
 but also because,
if confirmed on a larger sample of clusters at higher redshift, would imply
that the deviations from the hydrostatic equilibrium in the ICM are small,
and therefore that X-ray derived
galaxy cluster masses using this assumption are
reliable probes of cosmological parameters (\cite{allen11}).

Another important result came from the study of the chemistry of the ICM
in the Perseus cluster. The unprecedented combination of resolving power
and effective area allowed the {\it Hitomi}/SXS to measure the
relative elemental abundance of iron-peak elements such as chromium,
manganese, and nickel (cf. the {\it right panel} of Fig.~\ref{fig1}).
{\it Hitomi} demonstrated that these abundances
are consistent with solar, disproving prior claims of significant
over-abundance based on CCD-resolution data (\cite{hitomi17a}). This
constraints the progenitor of type Ia SN to be a combination of
near- and sub-Chandrasekhar.

Additional papers published on the SXS observation of the Perseus cluster
discuss: resonant scattering in the ICM core
(\cite{hitomi18a}), tight constraints on the
X-ray decay signature of sterile neutrinos, a possible dark matter candidate
(\cite{hitomi17b});
the bulk velocity field (\cite{hitomi18b})
and the temperature structure of the ICM (\cite{hitomi18c}).
Readers are referred to
the contribution by Tamura in these Proceedings
for a full description of the whole range of scientific results of the
{\it Hitomi} observations of the Perseus Cluster.

\section{The Hot Universe with XRISM}

In the light of the extraordinary results obtained by {\it Hitomi} during its,
alas, too short!, operational life, JAXA and NASA decided to propose
a mission to recover one of its fundamental
scientific objectives: ``Resolving
astrophysical problems by precise high-resolution X-ray spectroscopy''. This
is the {\bf X-Ray Imaging and Spectroscopy Mission} (XRISM, a.k.a XARM)
(\cite{tashiro18}). The European
and Canadian Space Agencies, as well as European institutes participate
in the mission development together with a wide range of scientific institutions
in Japan and the United States. The mission is designed to fulfill the main
theme: ``Revealing material circulation and energy transfer in cosmic
plasma and elucidating evolution of cosmic structure and objects'' with
a payload that closely replicates the soft X-ray telescopes and instruments
on {\it Hitomi}:

\begin{itemize}

\item a micro-calorimeter ({\it Resolve}) with
  a requirement energy resolution $\le$7~eV in the 0.3-12~keV
  energy range over a 3'$\times$3' field-of-view covered by an array of
  35 sky sensitive pixels

\item an array of CCD detector ({\it Xtend}) with a large field-of-view
  (larger than 30'$\times$30'), and an energy resolution $\le$200~eV at 6~keV
  at the beginning of
  the operational life

\item a large-area, light weight soft X-ray telescope with a $\simeq$1.7'
  Half Energy Width (HEW) or better, and an area comparable to that of the soft X-ray
  telescopes on-board {\it Hitomi}
  
\end{itemize}

\begin{figure}[b]
  \begin{center}
  \includegraphics[width=2.5in]{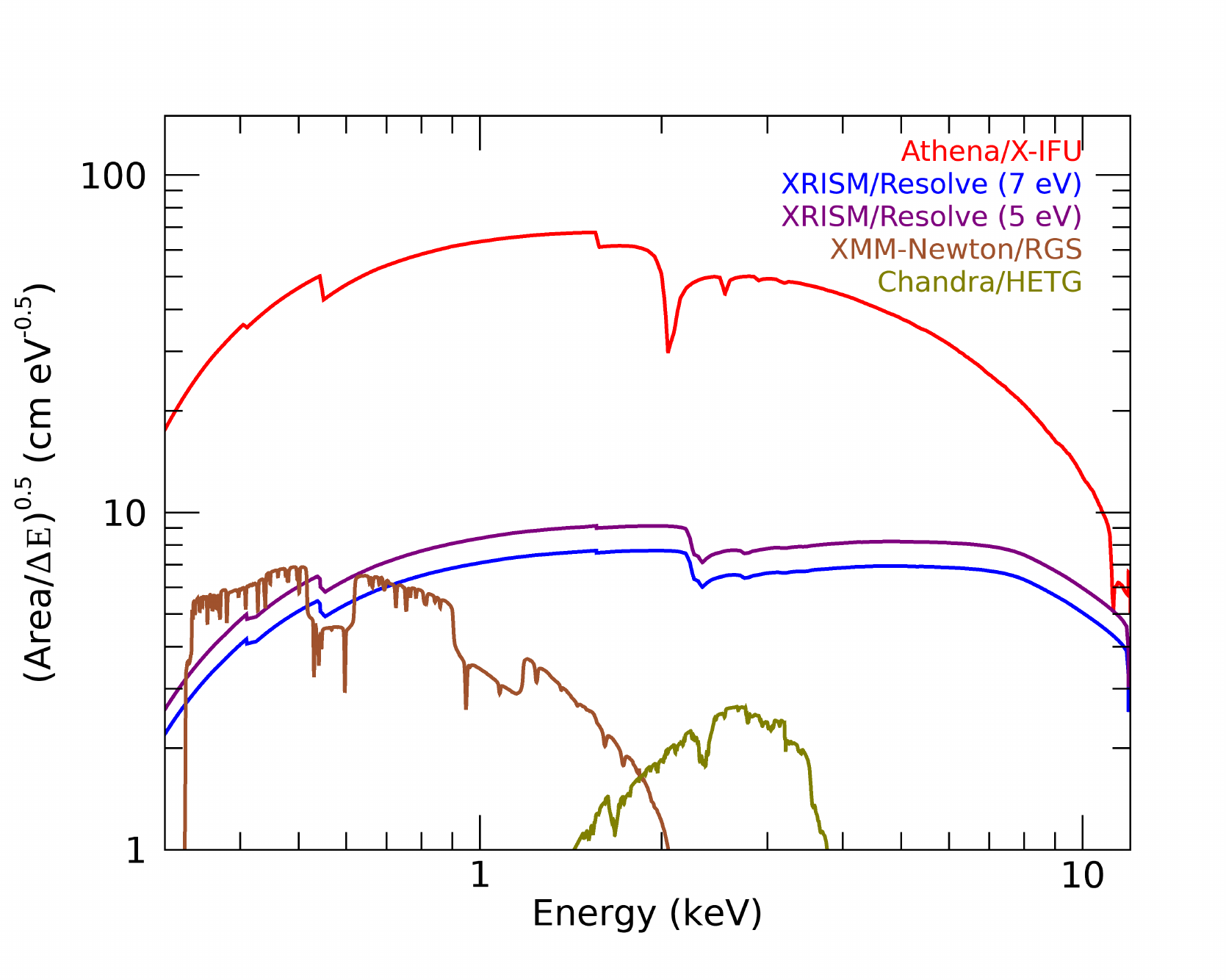}
  \includegraphics[width=2.5in]{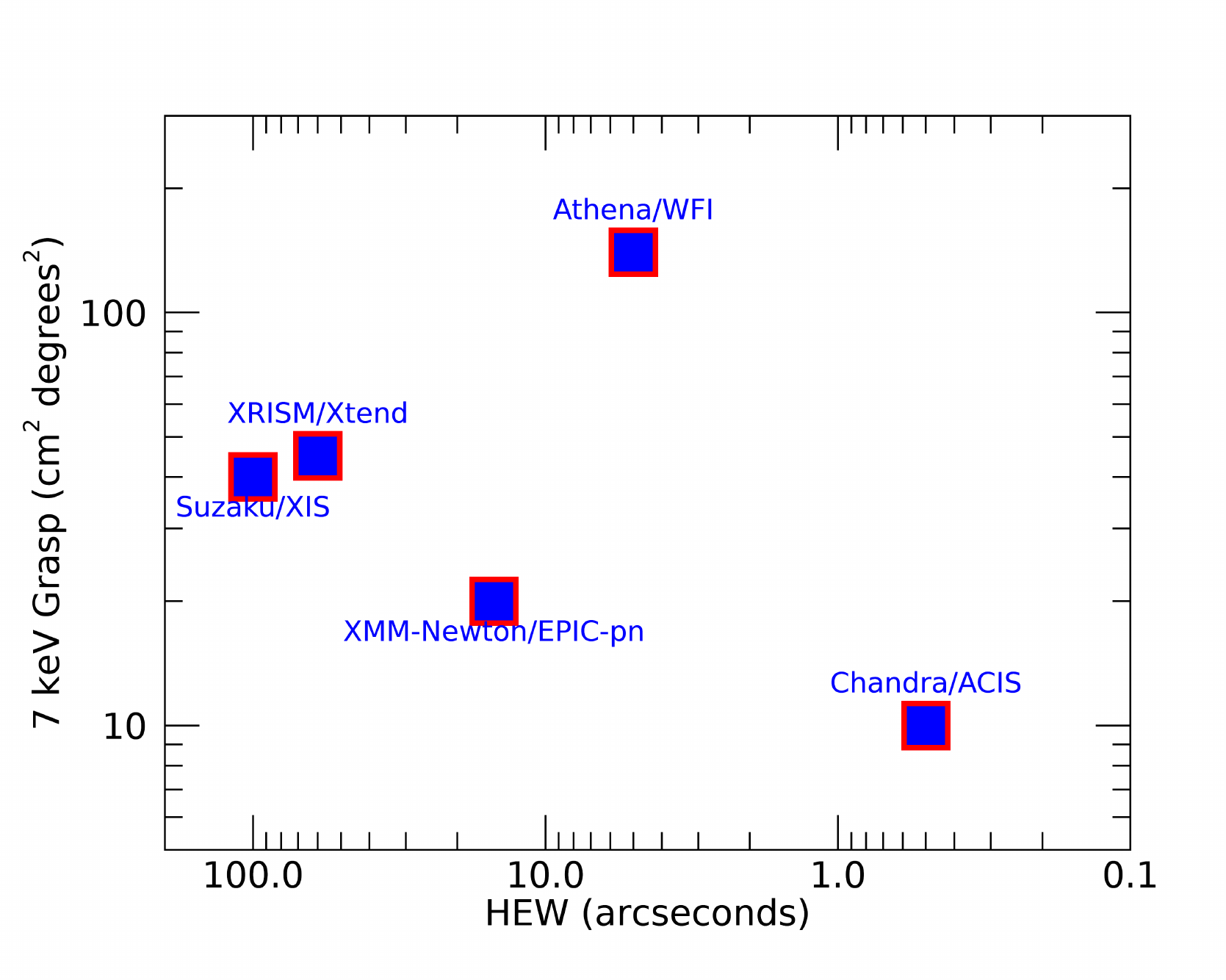}
  \caption{{\it Left panel:} Weak-line X-ray spectroscopy figure-of-merit
    for
        selected
        operational and future X-ray observatories. The figure of merits is the square
        root of the ratio between the effective area and the energy resolution. For
        the {\it Resolve} instrument on-board XRISM two values are shown, based on
        the energy resolution requirements (7~eV) and the proven flight resolution
        of the {\it Hitomi} SXS ($\le$5~eV). {\it Right panel}: 7~keV Grasp versus HEW
        for selected operational and future X-ray observatories. The 1~keV grasp, where
        the {\it Athena}/SPO area is optimized, is $\simeq$2800~cm$^2$ degrees$^2$ for the
        {\it Athena}/WFI, $\simeq$400~cm$^2$ degrees$^2$ for the EPIC-pn, and
        $\simeq$50~cm$^2$ degrees$^2$ for the {\it Chandra}/ACIS.
        }
  \label{fig2}
  \end{center}
\end{figure}

This configures a payload whose X-ray spectroscopic performance largely exceeds
any currently operational mission (Fig.~\ref{fig2}).

One of the XRISM Science Objective is directly related to the ``Hot Universe'':
``Structure, formation and evolution of cluster of galaxies''. A
micro-calorimeter like {\it Resolve} will be able to address themes that
have been only initiated by {\it Hitomi}. Elucidating how the gravitational
energy released by cluster mergers is converted into thermal energies, and
how the energy is distributed between particle and 
collective gas motions is one of the main goals of XRISM. These
objectives will be pursued through measurements of the velocity field
structure in the central regions of cool clusters to examine local heating
sources (AGN jets; magneto-hydrodynamic interaction between the ICM
and member galaxies); measurements of temperature and collective motions
of gas stripped from galaxy group accreting onto a cluster to investigate
if the infalling galaxies contribute to the ICM heating; and measurements of the
turbulent velocity in relaxed and disturbed galaxies that will allow us to
evaluate how gravitational energy is distributed among thermal energy,
kinetic motions of the ICM, and relativistic particles. Not less
importantly in a cosmological context, extending the sample of
measurements of the ICM turbulent pressure will allow us
to correct the hydrostatic bias potentially affecting the X-ray
cluster mass
functions and therefore remove systematics in the determination
of cosmological parameters. XRISM
will continue the investigation on the metallicity of the gas
trapped in the filament of the cosmic web as a probe of the
contribution of different SN explosion types and progenitor populations to
the cosmic nucleosynthesis.
XRISM is due to launch in the early 2020s (\cite{tashiro18}).

\section{The Hot Universe with Athena}

The next step in this challenge is {\it Athena}, the second L-class mission of the
ESA ``Cosmic Vision'' program. {\it Athena} (\cite{nandra13})
is a large area observatory, aiming
at addressing the science themes of the ``Hot and Energetic Universe''.
{\it Athena} aims at tracing the chemical and physical evolution of
large-scale cosmic structures from the epoch of their formation (z$\sim$2--3) to
the present Universe; and to study the evolution of accreting black holes in the
Universe and of the processes through which they affect the cosmological evolution of
the galaxy where they reside, by performing a full census of AGN up to the
epoch of reionization. However, besides these basic core science themes, {\it Athena} will
be an observatory fully open to the international astronomical community, with
fast ($\le$4~hours) and efficient ($\simeq$50\%)
response to
Targets of Opportunity occurring in a random position in the sky.
The large majority of its observing time will be allocated on the basis of proposals
evaluated in a peer review process.
This high-level scientific objectives will be achieved through an innovative
payload:

\begin{itemize}
  
\item a single telescope based on Silicon Pore Optics (SPO) technology developed
  in Europe (\cite{collon16}), with a 12~m focal length, 5'' HEW
  angular resolution at energies lower than 7~keV, and an effective area
  $\ge$1.4~m$^2$ (0.25~m$^2$) at 1 (6)~keV. 
  
\item the Wide Field Imager (WFI; \cite{meidenger16}), an active pixel sensor Si detector
  with
  a wide field-of-view (40'$\times$40'), and spectral-imaging capability with
  a CCD-like energy resolution ($\le$150~eV at 6~keV)

\item the X-ray Integral Field Unit (X-IFU; \cite{barret18}) a
  cryogenic imaging spectrometer with a 2.5~eV energy resolution in the
  energy range between 0.2 and 12~keV, 5' diameter effective field-of-view
  and a $\simeq$5'' pixel size

\end{itemize}

This payload configuration guarantees that the {\it Athena} performance will exceed
any existing of planned X-ray observatory in
the decade of the 20s by more than one order-of-magnitude in several parameter spaces
simultaneously, such as the photon collecting area
or the truly unprecedented combination of spatial resolution,
energy resolution and field-of-view enabled by the X-IFU (Fig.~\ref{fig2}).

\subsection{Formation and evolution of galaxy group and clusters}

{\it Athena} will probe for the first time the physical properties of the ICM in
galaxy clusters and group at the epoch of their formation (z~$\simeq$~2). Measurement
of X-ray surface brightness and temperature with the X-IFU and WFI will allow to
derive the entropy profile, and compare it with the expectations of a pure gravitational
collapse as opposed to external effects such as feedback by SN winds and/or AGN.
These accurate measurements will permit tracing the evolution of scaling relations
between the total mass of the cluster with X-ray observables such as the
temperature, X-ray luminosity, entropy up to z$\simeq$2 for a wide range of masses
(\cite{pointecouteau13}),
thus ideally complementing the full census of z$\le$1.5 massive clusters to be
achieved by eROSITA (\cite{merloni12}). In the local Universe, the X-IFU will
map the non-thermal component of the ICM energy budget through spatially-resolved
spectroscopy of the line spectral broadening due to turbulence, and of the line
shift due to bulk motion, determining for the first time the large scale properties
of the ICM for nearby clusters by resolving the relevant spatial scales.
Maps of emission measure, temperature and metallicity in the
outskirts of local clusters will unveil the processes occurring where new material
is accreting into the dark matter potential and energy is transferred to the ICM
via merging events (\cite{ettori13}).

\subsection{Chemical history of hot baryons}

High-resolution spectroscopy of the ICM X-ray line-rich spectrum is the only way
to probe the metal abundances of the gas and its cosmological evolution. In the local
Universe,
{\it Athena} will map the distribution of the most abundant elements out to the
outskirts of galaxy clusters. The peripheral abundance will be used to
estimate the contribution of AGN feedback in expelling pre-enriched
gas. {\it Athena} will measure the metallicity of
a wide range of elements up to z$\simeq$2, with accuracies ranging a a few percent
(Fe, Si) to 10--20\% (O, Ca) (\cite{pointecouteau13}).
Because the ICM is an almost perfectly isolated system
in the dark matter potential well, its abundances are the
resultant of the metal synthesis by different types of SN over cosmic
time (a measurement pioneered by {\it Hitomi} on the Perseus Cluster: \cite{hitomi17a}).
Ratios of iron-peak elements such as chromium and nickel against iron trace SN Ia,
those of intermediate-Z elements from
O to Si trace core-collapsed SN, lower Z elements
are mostly sensitive to AGB stars (\cite{werner08}). The {\it
Athena} X-IFU will measure the whole range of these ``metals'', and provide
a full characterization of the element synthesis history
(\cite{ettori13}).

\subsection{AGN feedback in galaxy groups and clusters}

The {\it Athena} X-IFU will
measure velocities of the hot ICM gas with a precision of 10-20~km~s$^{-1}$,
and temperatures and
metallicity with a precision of a few percent on spatial scales of $\simeq$5''
in nearby cluster cores.
This implies determining the kinematics of the hot gas in galaxies, groups and
clusters on scales small enough to be able to resolve the regions where
powerful AGN jets impact, and potential affect its dynamical state. This will
allow us to assess how the energy carried by relativistic particles is
dissipated and distributed to the ICM, and how the balance between
heating and cooling is
maintained in regions where the most massive galaxies are formed. In other
terms, while we have now a good understanding of the macro-physics of the
radio mode of AGN feedback in nearby clusters (\cite{hlavaceklarrondo15}),
{\it Athena} will investigate
for the first time the micro-physics of this process thanks to X-IFU spatially-resolved
spectroscopy of the hot gas. By estimating the energy stored in cavities,
ripples and motions, {\it Athena} will estimate the energy input of AGN for
the first time. On a larger (Mpc) scale,
velocity measurements of shocked expanding hot shells around radio lobes
will allow an estimate of the integrated
jet power, an elusive quantity preventing the whole power of strong shocks from being
estimated (\cite{croston13}).
Furthermore, the
WFI large field-of-view
(coupled with the smooth vignetting curve of the SPO, \cite{willingale13})
will enable population studies of AGN-induced disturbances and bow shocks
surrounding radio lobes, permitting
to characterize morphological disturbances of the surface brightness distribution
in tens of clusters at z$\le$0.1, and to
correlate their mechanical energy with the properties of the environment and
the AGN jet power (\cite{croston13}).

\subsection{Missing baryons and WHIM}

About one half of the baryons in the Universe are still elusive.
Cosmological simulations shows that
the bulk of them should be locked in a hot medium, the Warm-Hot Intergalactic
Medium (WHIM), with temperature ranging between 10$^5$ to 10$^7$~K. However,
they have escaped detection so far (see, however, \cite{nicastro18} for a
different view). This cosmic ``hide-and-seek'' will end with {\it Athena} that
will be able to detect the WHIM, and characterize its physical properties,
both in absorption against a bright background source (Gamma-Ray Bursts, GRB;
blazars,
etc.), or through emissions along tenuous and diffuse filaments. During its
nominal 4~year operation life, {\it Athena} is expected to observe about
90 systems in absorption, follow-up $\simeq$7 GRB in emission, and observe
about 100 shallow and 1 deep side-line in emission. Depending of the
speed of the spacecraft re-pointing, this will allow to probe filament 
with Oxygen column densities down to 10$^{15-16}$~cm$^{-2}$ (\cite{kaastra13}).

\section{Summary}

After the global disarray for the early demise of {\it Hitomi} operations, the
future of X-ray astronomy looks bright again thanks to a plethora of new
missions and observatories due launch in the next few years. In this
context, the enhanced high-resolution capabilities of the XRISM/{\it Resolve},
enhanced in the {\it Athena}/X-IFU
by its sharp angular resolution and better energy resolution, and combined 
to deep wide-field spectral imaging with the XRISM/{\it Extend} and the
{\it Athena}/WFI will yield a giant leap in our understanding of the formation and
evolution of large-scale baryonic structure in the ``Hot Universe''.
Furthermore, besides
their nominal core science goals XRISM and {\it Athena} are observatories open
to the astronomical community. Their observational programs
will be primarily driven by
the collective scientific wisdom of the community active at the time of their
operations. Thanks to the
unique capabilities of XRISM and {\it Athena} - largely exceeding
those of any existing or planned spectroscopic X-ray missions -
we should ``expect the unexpected'' - as the far too short {\it Hitomi} history
clearly shows. 

\section*{Acknowledgments}

This paper is written on behalf of the XRISM Science Management Office and of
the Athena Science Study Team, whose comments on an early version of the
manuscript are gratefully acknowledged. We thank Dr. Didier Barret for
a careful revision of the manuscript.


\begin{thebibliography}{}

\bibitem[Allen \etal\ 2011]{allen11}
{Allen, S.W., Evrard, A.E., Mantz, A.B.} 2011, 
\textit{ARA\&A}, 49, 409

\bibitem[Barret \etal\ 2016]{barret16}
{Barret, D., \etal\ } 2016, 
\textit{SPIE}, 9905, 2

\bibitem[Barret \etal\ 2018]{barret18}
{Barret, D., \etal\ } 2018, 
\textit{SPIE}, in press (arXiv:1807.06092)

\bibitem[B\"ohringer \etal\ 1993]{boehringer93}
{B\"ohringer, H., Voges, W., Fabian, A.C., Edge, A.C., Neumann, D.M. } 1993, 
\textit{MNRAS}, 264, 25

\bibitem[Brinkman \etal\ 1987]{brinkman87}
{Brinkman, A.C., \etal\ } 1987, 
\textit{ApL\&C}, 26, 73

\bibitem[Canizares \etal\ 2005]{canizares05}
{Canizares, C.R., \etal\ } 2005, 
\textit{PASP}, 117, 1144

\bibitem[Collon \etal\ 2016]{collon16}
{Collon, M., \etal\ } 2016, 
\textit{SPIE}, 9905, 28

\bibitem[Croston \etal\ 2013]{croston13}
{Croston, J.H., \etal\ } 2013, 
\textit{arXiv}, 1306.2323

\bibitem[den Herder \etal\ 2001]{denherder01}
{den Herder, J.W., \etal\ } 2001, 
\textit{A\&A}, 365, 7

\bibitem[Ettori \etal\ 2013]{ettori13}
{Ettori, S., \etal\ } 2013, 
\textit{arXiv}, 1306.2322

\bibitem[Fabian \etal\ 2000]{fabian00}
{Fabian, A.C. \etal\ } 2000, 
\textit{MNRAS}, 318, 65

\bibitem[Fabian 2012]{fabian12}
{Fabian, A.C.} 2012, 
\textit{ARA\&A}, 50, 455

\bibitem[Hitomi Collaboration 2016]{hitomi16}
{Hitomi Collaboration} 2016, 
\textit{Nature}, 535, 117

\bibitem[Hitomi Collaboration 2017a]{hitomi17a}
{Hitomi Collaboration} 2017a, 
\textit{Nature}, 551, 478

\bibitem[Hitomi Collaboration 2017b]{hitomi17b}
{Hitomi Collaboration} 2017b,
\textit{ApJ}, 837, 15

\bibitem[Hitomi Collaboration 2018a]{hitomi18a}
{Hitomi Collaboration} 2018a, 
\textit{PASJ}, 70, 10

\bibitem[Hitomi Collaboration 2018b]{hitomi18b}
{Hitomi Collaboration} 2018b, 
\textit{PASJ}, 70, 9

\bibitem[Hitomi Collaboration 2018c]{hitomi18c}
{Hitomi Collaboration} 2018c, 
\textit{PASJ}, 70, 11

\bibitem[Hlavacek-Larrondo \etal\ 2015]{hlavaceklarrondo15}
{Hlavacek-Larrondo, J.; \etal\ } 2015
\textit{ApJ}, 805, 35

\bibitem[Kaastra \etal\ 2013]{kaastra13}
{Kaastra, J., \etal\ } 2013, 
\textit{arXiv}, 1306.2324

\bibitem[Kelley \etal\ 2016]{kelley16}
{Kelley, R.L., \etal\ } 2016
\textit{SPIE}, 9905, id.~99050V

\bibitem[JAXA 2016]{jaxa16}
{JAXA} 2016
\textit{Hitomi Experience Report: Investigation of Anomalies Affecting the X-ray Astronomy Satellite 'Hitomi' (ASTRO-H)}

\bibitem[Meidenger \etal\ 2016]{meidenger16}
{Meidinger, N., \etal\ } 2016, 
\textit{SPIE}, 9905, 2

\bibitem[Merloni \etal\ 2012]{merloni12}
{Merloni, A., \etal\ } 2012, 
\textit{arXiv}, 1209.3114

\bibitem[Nandra \etal\ 2013]{nandra13}
{Nandra, K., \etal\ } 2013, 
\textit{arXiv}, 1306.2307

\bibitem[Nicastro \etal\ 2017]{nicastro17}
{Nicastro, F., Krongold, T., Mathur, S., Elvis, M.} 2017, 
\textit{AN}, 338, 218

\bibitem[Nicastro \etal\ 2018]{nicastro18}
{Nicastro, F. \etal\ } 2018, 
\textit{arXiv}, 1806.08395

\bibitem[Paerels \& Kahn 2003]{paerels03}
{Paerels, F.B.S., Kahn S.M.} 2003, 
\textit{ARA\&A}, 41, 291

\bibitem[Pointecouteau \etal\ 2013]{pointecouteau13}
{Pointecouteau, E., \etal\ } 2013, 
\textit{arXiv}, 1306.2319

\bibitem[Schaye \etal\ 2015]{schaye15}
{Schaye, J., \etal\ } 2015, 
\textit{MNRAS}, 446, 521

\bibitem[Tashiro \etal\ 2018]{tashiro18}
{Tashiro, M., \etal\ } 2018, 
\textit{SPIE}, in press

\bibitem[Tsujimoto \etal\ 2018]{tsujimoto18}
{Tsujimoto, M., \etal\ } 2018, 
\textit{JATIS}, 4, 1205

\bibitem[Werner \etal\ 2008]{werner08}
{Werner, N.; Finoguenov, A.; Kaastra, J. S.; Simionescu, A.; Dietrich, J. P.; Vink, J.; B\"ohringer, H.} 2018, 
\textit{SSRv}, 134, 337

\bibitem[Willingale \etal\ 2013]{willingale13}
{Willingale, R., \etal\ } 2013, 
\textit{arXiv}, 1307.1709

\end{thebibliography}
\end{document}